\documentclass[aps,pre,reprint]{revtex4-1}
\usepackage{amsmath}
\usepackage[latin1]{inputenc}
\usepackage{graphicx}

\usepackage[]{hyperref}
\usepackage{bm}

\usepackage{epsfig}

\usepackage{color}
\usepackage{xcolor}
\hypersetup{
    colorlinks,
    linkcolor={red!50!black},
    citecolor={blue!50!black},
    urlcolor={blue!80!black}
}

\begin{document}
\title{The role of sample height in the stacking diagram of colloidal mixtures under gravity}

\author{Thomas Geigenfeind}
\affiliation{Theoretische Physik II, Physikalisches Institut,
 Universit{\"a}t Bayreuth, D-95440 Bayreuth, Germany}

\author{Daniel de las Heras}
\email{delasheras.daniel@gmail.com}
\homepage{www.danieldelasheras.com}
\affiliation{Theoretische Physik II, Physikalisches Institut,
 Universit{\"a}t Bayreuth, D-95440 Bayreuth, Germany}

\date{\today}
\begin{abstract}
Bulk phase separation is responsible for the occurrence of stacks of different layers in sedimentation of colloidal mixtures. 
A recently proposed theory (de las Heras and Schmidt 2013 Soft Matter 9 8636) establishes a unique connection between
the bulk phase behaviour and sedimentation-diffusion-equilibrium. The theory constructs a stacking diagram of all possible
sequences of stacks under gravity in the limit of very high (infinite) sample heights. Here, we study the stacking diagrams of colloidal mixtures at 
finite sample height, $h$. We demonstrate that $h$ plays a vital role in 
sedimentation-diffusion-equilibrium of colloidal mixtures. The region of the stacking diagram occupied
by a given sequence of stacks depends on $h$. Hence, two samples with different heights but identical colloidal
concentrations can develop different stacking sequences. In addition, the stacking diagrams for different heights can be qualitatively different
since some stacking sequences occur only in a given interval of sample heights.
We use the theory to investigate the stacking diagrams of both model bulk systems and mixtures of patchy particles that differ 
either by the number or by the types of patches. 

\end{abstract}

\maketitle

\section{Introduction \label{introduction}}

Since the pioneer work of Perrin~\cite{perrin}, sedimentation has become a central tool for investigating the phase behaviour in colloidal systems.
The height-dependent colloidal concentration profile provides a direct measurement of the equation of state for monocomponent systems~\cite{PhysRevLett.71.4267,biben1993density}.
Sedimentation experiments are also used to extract information from the bulk phase behaviour in binary colloidal mixtures, see e.g. Refs.~\cite{Piazzareview,C1SM06535A,C5SM00615E}. 
However, thermal and gravitational energies are of the same order of magnitude for typical colloidal systems. This results in additional gravity-induced phenomenology not present in bulk systems.
Examples are denser particles floating on top of lighter colloids~\cite{piazza1}, a nematic layer sandwiched by isotropic layers in mixtures of platelets and spheres~\cite{floating}
and mixtures of thin and thick rods~\cite{C6SM00736H}, and reentrant network formation in mixtures of patchy colloids~\cite{Lucas}. It is also common to observe complex stacking sequences 
in sedimentation with three or more different layers, such as e.g., in mixtures of charged platelets and polymers~\cite{doi:10.1021/la804023b}, plate-plate binary systems~\cite{C3SM52311J},
mixtures of spheres of different sizes~\cite{ahhh}, and colloidal rod-plate mixtures~\cite{JeR}.

The relation between bulk phase behaviour and sedimentation-diffusion-equilibrium in mixtures is therefore
intertwined with gravity-induced effects. From a theoretical viewpoint,  
a generalized Archimedes principle was formulated~\cite{piazza1,piazza2}
for the case where one of the components is very diluted. Sedimentation was also studied
by analyzing the macroscopic osmotic equilibrium conditions~\cite{0295-5075-66-1-125,PhysRevE.70.051401}.
Recently, de las Heras and Schmidt have proposed a theory~\cite{stack1,stack2} for obtaining the stacking diagram, i.e.,
the set of all possible stacking sequences under gravity, from the bulk phase diagram of a given binary system.
The theory is based on the concept of sedimentation paths. Each sedimentation path, which is a line in the plane of chemical potentials representation of the bulk phase diagram,
describes the state of the mixture under gravity, in sedimentation-diffusion-equilibrium. Using this theory the stacking
diagrams of mixtures of spheres and platelets~\cite{stack1} and mixtures of platelets and nonadsorbing polymers~\cite{stack2} were obtained. Also, very recently,
van Roij and coworkers have obtained the stacking diagrams of mixtures of thin and thick colloidal rods~\cite{C6SM00736H}. Although in all these cases the bulk
phase diagrams of the colloidal systems are relatively simple, the resulting stacking diagrams are extremely rich and show complex stacking sequences.
These works are focused on the limit of very high (infinite) sample heights. This idealized limit is very relevant in experimental work since the height of the test tube
is typically larger than the gravitational lengths of the colloids.

The interplay between micro confinement and colloidal sedimentation has been experimentally and theoretically investigated~\cite{Royall}. However,
little attention has been paid to the influence of the total (macroscopic) sample height in colloidal sedimentation. A remarkable exception is the experimental work of
Jamie et al.~\cite{B915788C}, in which the properties of the gas-fluid interface of a polymer-colloid mixture are analyzed as a function of the overall height of the container.
By systematically changing the total sample height while keeping the polymer-colloid concentrations fixed, the interfacial properties were found to move towards the critical point.
Theoretically, it has been shown that varying the sample height might led to a change in the stacking sequence in mixtures of colloids and nonadsorbing polymers~\cite{0295-5075-66-1-125,stack2}. 

Here, we use the theory of Refs.~\cite{stack1,stack2} to study sedimentation-diffusion-equilibrium of colloidal mixtures for the case of finite height samples.
We systematically investigate the role of sample height in the stacking diagrams of colloidal mixtures. 
We first apply the theory to model binary systems. That is, systems with generic bulk phase diagrams typical of model Hamiltonian which we however do no explicitly specify. 
As an application of current interest we also study sedimentation in patchy colloidal mixtures. Patchy colloids are functionalized colloids that interact via a directional and valence-limited
potential~\cite{in1,in2}. Here, we use Wertheim's theory~\cite{wertheim} to obtain the bulk behaviour of the mixture. The two species of the mixture differ in either the number or in the types of patches. 
We show that the sample height is a crucial variable in sedimentation-diffusion-equilibrium of colloidal mixtures. The stacking diagrams for the same mixture but for
different sample heights differ not only quantitatively but also qualitatively. For example, some stacking sequences occur only in a given range of sample heights.

\section{Theory \label{theory}}

\subsection{The sedimentation path}
Consider a colloidal mixture under gravity in a sedimentation vessel of sample height $h$. The gravitational potential
for each species $i=1,2$ is  $m_igz$, where $m_i$ is the buoyant mass of species $i$, $g$ is the acceleration
due to gravity, and $z$ is the vertical coordinate (we set the origin of coordinates, $z=0$, at the bottom of 
the sample). Using a local density approximation~\cite{schmidt04aog,floating,stack1,stack2}, we define a height-dependent local chemical potential for $0\le z\le h$ for each 
species
\begin{eqnarray}
\mu_1(z)=\mu_1^{\text{b}}-m_1gz,\nonumber\\
\mu_2(z)=\mu_2^{\text{b}}-m_2gz,\label{chemi}
\end{eqnarray}
where $\mu_i^{\text{b}}$ is the bulk chemical potential, i.e. the chemical potential in absence of gravity.
The local density approximation assumes that for each $z$ the state of the sample is analogous to 
a bulk system (no gravity) with chemical potentials given by~\eqref{chemi}. This constitutes a very good approximation
provided that the correlation lengths are small compared to the gravitational lengths, $\xi_i=k_{\text{B}}T/m_ig$ with 
$k_{\text{B}}$ the Boltzmann constant, and $T$ the absolute temperature. Combining the expressions for the local chemical
potentials, cf.~\eqref{chemi}, and eliminating the height variable $z$ we find
\begin{equation}
\mu_2(\mu_1)=a+s\mu_1,\label{path}
\end{equation}
where both $a$ and $s$ are constants given by
\begin{eqnarray}
a&=&\mu_2^b-s\mu_1^b,\nonumber\\
s&=&m_2/m_1=\xi_1/\xi_2.\label{constants}
\end{eqnarray}
The finite size of the sample $0\le z\le h$ is translated into a range for the local chemical potentials
\begin{equation}
0\le\frac{\mu_i^{\text{b}}-\mu_i}{m_igh}\le1,\text{ }i=1,2.~\label{limit}.
\end{equation}
Eqs.~\eqref{path} and~\eqref{limit} represent a line segment, which we refer to as the sedimentation path, in the plane of chemical potentials.
The sedimentation path describes how the local chemical potentials vary along the height coordinate in the vessel.
Each point in the sedimentation path corresponds to the state of the sample at a given $z$. 

\begin{figure}
\epsfig{file=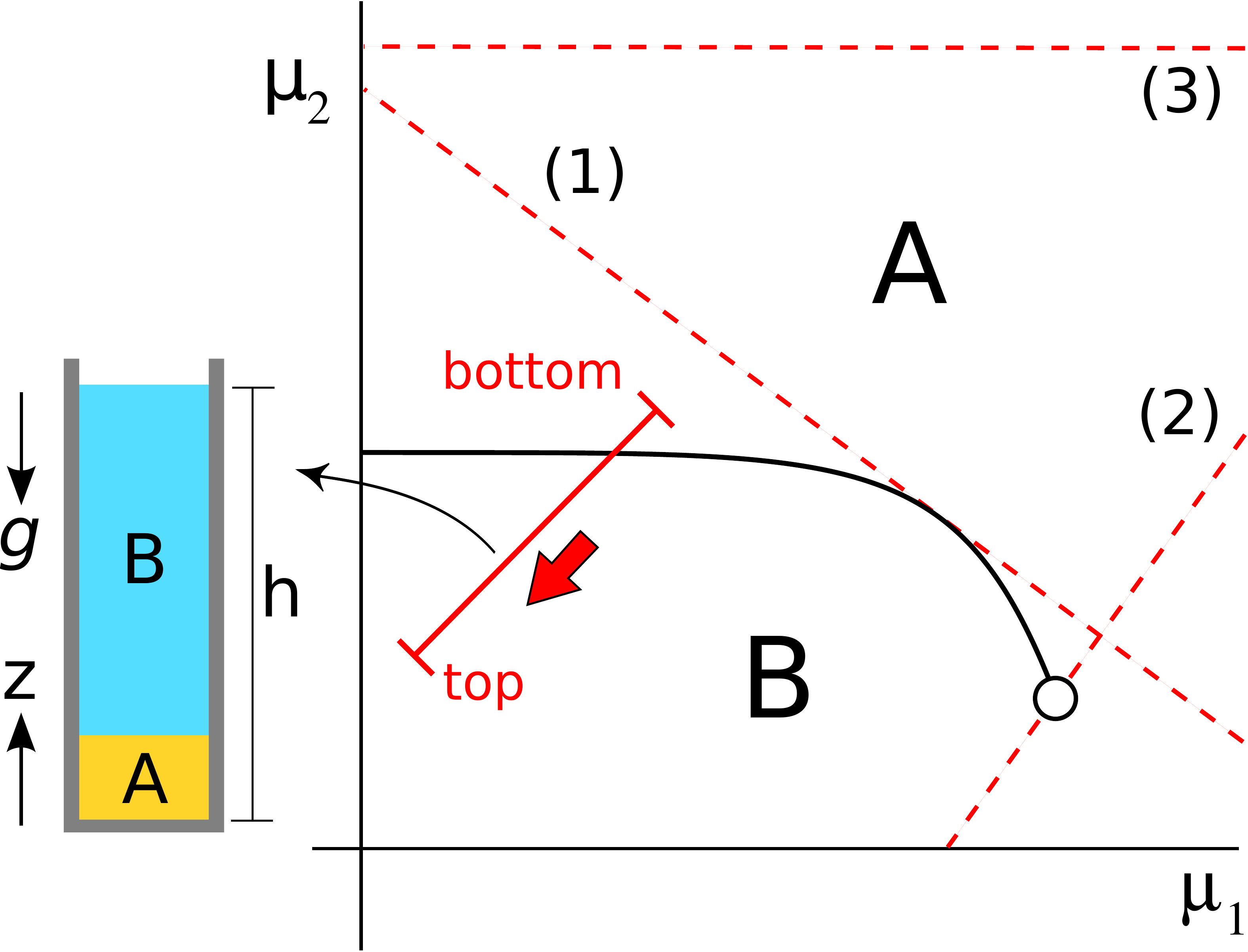,width=0.8\columnwidth}
\caption{Bulk phase diagram (schematic) of a binary mixture in the plane of chemical potentials $\mu_1-\mu_2$. Two phases A and B coexist at the binodal (solid black line).
The binodal ends at 
a critical point (empty circle) and has a horizontal asymptote since the pure system of species $2$ undergoes a phase transition.
The solid red line represents the sedimentation path of the mixture in a vessel of height $h$ under gravity. The arrow indicates the direction of the path
from the bottom to the top of the sample. The corresponding stacking sequence is bottom A and top B, as shown in the sketch. The dashed red lines are selected
sedimentation paths for infinite height: (1) a path tangent to the binodal, (2) a path that crosses an ending point of a binodal, (3) a path parallel to
the asymptotic behaviour of the binodal.
} 

\label{fig1}
\end{figure}

The sedimentation path is directly related to the stacking sequence, i.e., the sequence of stacks of different phases that appear under gravity. 
If a path crosses a boundary between two phases in the phase diagram, e.g., a binodal, an interface appears in the vessel. 
The sedimentation path provides a direct link between the bulk phase diagram of the mixture and the stacking sequence.
An example of a sedimentation path and its corresponding stacking sequence is shown in Fig.~\ref{fig1}. 

A sedimentation path is fully described by its (i) slope, (ii) location in the $\mu_1-\mu_2$ plane specified by a point on the path, 
(iii) direction and (iv) length. The slope is fixed by  the ratio of the buoyant masses, cf.~\eqref{constants}. The position is determined by
the bulk chemical potentials in absence of gravity, and hence by the overall colloidal composition and concentration via a change of variables using
the equation of state of the mixture.
The direction is given by the signs of the buoyant masses (note $m_i$ can be negative if the mass density of the solvent
is higher than the mass density of the colloids). Finally, the length of the path is proportional to the height of the vessel
since the difference in chemical potentials between the top ($z=h$) and the bottom of the sample ($z=0$) is
\begin{equation}
\Delta\mu_i=\mu_i(h)-\mu_i(0)=-m_igh.\label{length}
\end{equation} 

\subsection{The stacking diagram}\label{theory}
We have shown how each sedimentation path is associated to a stacking sequence. The phase stacking diagram is the set of all possible stacking sequences for a given mixture.

{\bf Infinite height}. For standard colloidal particles in typical sedimentation vessels the length of the sedimentation path is of several $k_{\text{B}}T$ in the $\mu_1-\mu_2$ plane.
That is, the path extends over a big region of the bulk phase diagram of the mixture. Hence, a very relevant idealization is to consider the limit of very high (infinite) sample heights. 
Within this limit~\cite{stack1,stack2} a sedimentation path is a straight line of infinite length (not a line segment) in the plane
of chemical potentials. Hence, a sedimentation path can be fully described by 
using only the slope of the path $s$, and the crossing point between the path and the $\mu_2$ axis $a$, cf. \eqref{constants}. The stacking diagram can then be represented in the $s-a$ plane.
There are three types of boundaries between different stacking sequences in the stacking diagram. Here we only describe each one briefly, see~\cite{stack1,stack2} for a full account:

(i) {\it  Sedimentation binodals.} The set of all sedimentation paths tangent to a binodal in the bulk phase diagram is a boundary between two phases in the stacking diagram.
The path labeled as (1) in Fig.~\ref{fig1} is an example. An infinitesimally small change in one or in both variables of the path, $a$ and $s$, can change the stacking sequence. 

(ii) {\it Terminal lines.} The set of all paths crossing an ending point of a binodal in the bulk phase diagram is a boundary in the stacking diagram that 
we call the terminal line. The sedimentation path (2) in Fig.~\ref{fig1} is an example. An infinitesimal change of $a$ changes the stacking sequence.
An ending point can be e.g. a critical point, triple point, critical end point, etc. 

(iii) {\it Asymptotic terminal lines.} The third type of boundaries in the stacking diagram is formed by those paths that are parallel to the asymptotic behaviour of a binodal. See
the path (3) in Fig.~\ref{fig1}. In this case an infinitesimal change of the slope $s$ alters the stacking sequence.

A binodal is not the only possible boundary between two regions present in the bulk phase diagram. 
For example, a percolating line dividing the bulk phase diagram into percolated and nonpercolated states is another type of a boundary between phases.
Any boundary present in the bulk phase diagram generates boundaries in the stacking diagram (sedimentation binodals, terminal lines and asymptotic terminal lines). 
For convenience we speak always of binodal but one should bear in mind that other lines also give rise to boundaries in the stacking diagram. The patchy colloid mixtures
studied below feature such percolation lines.

{\bf Finite height}. In this paper we focus on the stacking diagrams for finite height samples. There exist several possibilities to represent the stacking diagram for finite heights.
In an experimental work one typically varies the concentration and composition of the mixture, while keeping the solvent and the mass density of the colloids unchanged. 
The sample height is, in principle, easy to adjust~\footnote{Solvent evaporation might occur, changing the effective sample height and hence the concentration of colloids.} and
hence forms a useful control parameter.
Under these circumstances the slope and the length of the path in the $\mu_1-\mu_2$ plane are fixed, cf.~\eqref{constants} and~\eqref{length},
and its position in the $\mu_1-\mu_2$ plane varies. A sensible choice of variables for the stacking 
diagram is the plane of average local chemical potentials along the path $\bar\mu_1-\bar\mu_2$. As the sedimentation paths are straight lines, the average local chemical potentials
are just the local chemical potential evaluated at the middle of the sample $\bar\mu_i=\mu_i(z=h/2)$. 

The stacking diagram for finite height samples in the $\bar\mu_1-\bar\mu_2$ plane contains three possible types of boundaries between different stacking sequences.
Two of them are sedimentation binodals originated from coexisting lines in the bulk phase diagrams and one boundary is due to the ending points of
the binodals. We describe each of them in detail in the following.

\begin{figure}
\epsfig{file=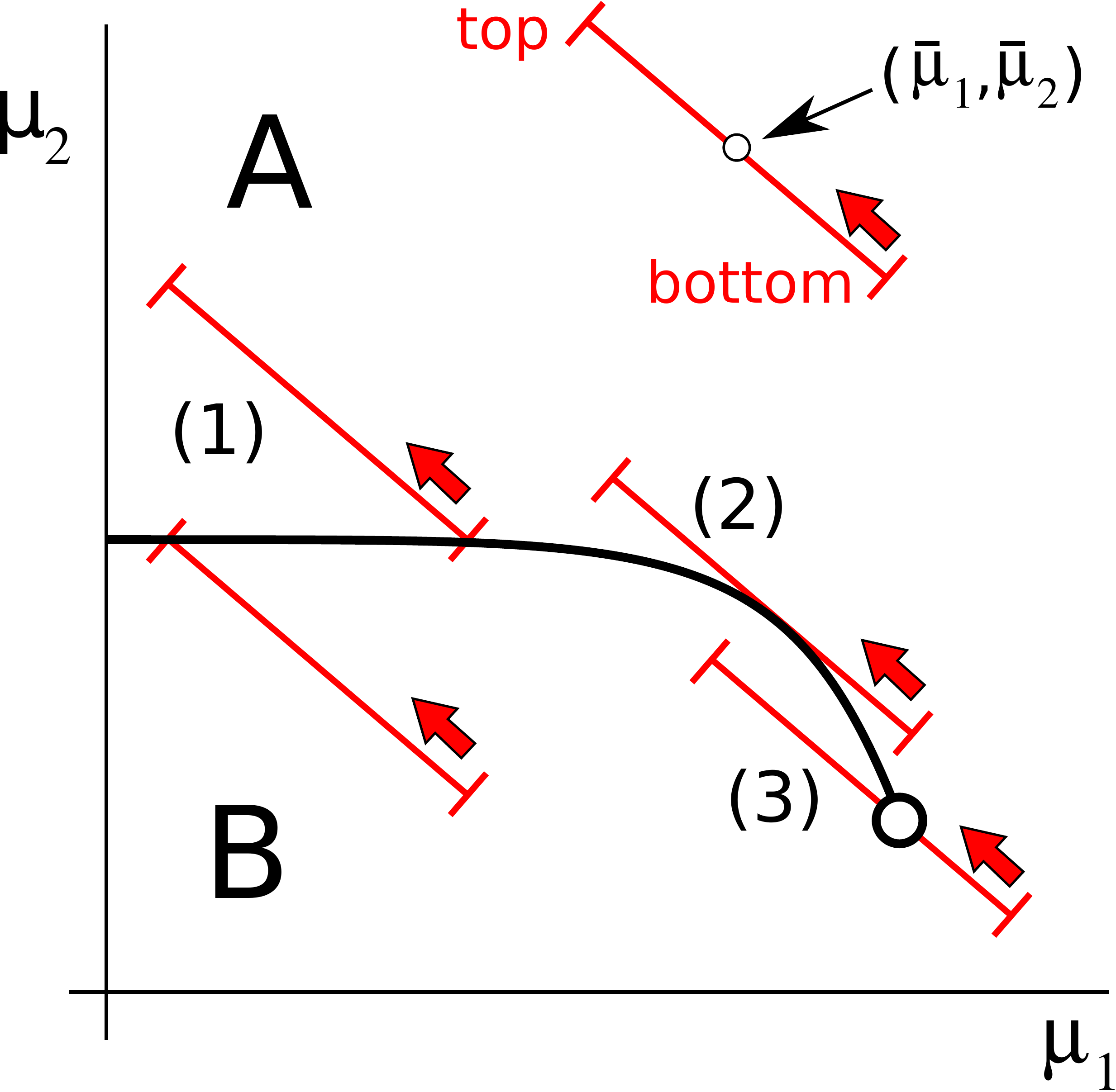,width=0.7\columnwidth}
\caption{Schematic bulk phase diagram of a binary mixture in the plane of chemical potentials $\mu_1-\mu_2$. Two phases A and B coexist at the binodal (solid black line). The binodal ends at 
a critical point (empty circle). The solid red lines are selected sedimentation paths (finite height). The arrow indicates the direction of the path
from the bottom to the top of the sample. The coordinates of the middle point of the path are the average local chemical potentials $(\bar\mu_1,\bar\mu_2)$ as indicated in one of the paths.
The top (bottom) path marked with (1) starts (ends) at the binodal. The path labeled as (2) is tangent to the binodal. The path (3) crosses the critical point.
} 
\label{fig2}
\end{figure}

(i) {\it Sedimentation binodal type I} (SBI). The set of all sedimentation paths that either start or end at a binodal form a sedimentation binodal of type I. 
A path starts (ends) at a binodal if the bottom (top) of the sample is located at the binodal. The path labeled as (1) in Fig.~\ref{fig2} is an
example. For each binodal in the bulk phase diagram there are two corresponding SBI in the stacking diagram. 
One SBI for those paths that end at the binodal and the other one SBI for those paths that start at the binodal. Both SBI lines have the same
shape as the bulk binodal. This type of boundary is not present in the case of infinite height because the paths do not have starting and ending points.

Let $\mu_{2,\text{C}}(\mu_1)$ be the parameterization of the chemical potential of species $2$ at a bulk coexistence line, such as a binodal, as a function of $\mu_1$.
Then, the two corresponding sedimentation binodals of type I are given by
\begin{eqnarray}
\bar\mu_2(\bar\mu_1)=\mu_{2,\text{C}}(\mu_{1}^-)+ m_2g\frac h2,\nonumber\\
\bar\mu_2(\bar\mu_1)=\mu_{2,\text{C}}(\mu_{1}^+)- m_2g\frac h2,
\end{eqnarray}
where 
\begin{equation}
\mu_1^{\pm}=\bar\mu_1\pm m_1g\frac h2.
\end{equation}
Here, $\mu_1^+$ ($\mu_1^-$) is the local chemical potential of species $1$ at the bottom (top) of the sample.

(ii) {\it Sedimentation binodal type II} (SBII). The set of all paths tangent to a bulk binodal is also a boundary (sedimentation binodal type II) 
in the stacking diagram. See the path (2) in Fig.~\ref{fig2}. This boundary is analogous to the sedimentation binodals in the case of infinite height.
The SBII boundaries are straight lines in the stacking diagram. A SBII line is present if and only if the slope of the path is the same as the
slope of the binodal at same point(s).  Each point of a binodal sharing the same slope as the path generates a SBII line.

Let $(\mu_{1,\text{t}},\mu_{2,\text{t}})$ be the chemical potentials of a point on a bulk binodal. Let its local slope be that of the sedimentation path. That is
\begin{equation}
\left.\frac{d\mu_{2,\text{C}}}{d\mu_1}\right|_{\mu_{2,\text{t}}}=s.\label{mmm}
\end{equation}
Then, the associated SBII line is given by
\begin{equation}
\bar\mu_2(\bar\mu_1)=\mu_{2,\text{t}}+(\bar\mu_1-\mu_{1,\text{t}})s.
\end{equation}
The finite size of the path limits the range of $\bar\mu_1$ to
\begin{equation}
\left|\bar\mu_1-\mu_{1,\text{t}}\right|\le\left|m_1g\frac h2\right|.
\end{equation}

(iii) {\it Terminal lines} (TL). The terminal lines are, as in the infinite height case, the set of all paths that cross an ending point of a binodal. See path (3) in Fig.~\ref{fig2}.
For each ending point in the bulk phase diagram there is one and only one TL in the stacking diagram. The TL is always a straight line. 

Let $(\mu_{1,\text{e}},\mu_{2,\text{e}})$ be the chemical potentials of an ending point in bulk, such as a critical point, a triple point, etc. The corresponding terminal line is 
\begin{equation}
\bar\mu_2(\bar\mu_1)=\mu_{2,\text{e}}+(\bar\mu_1-\mu_{1,\text{e}})s,
\end{equation}
for 
\begin{equation}
\left|\bar\mu_1-\mu_{1,\text{e}}\right|\le\left|m_1g\frac h2\right|.
\end{equation}

In the three cases (i),(ii), and (iii) any infinitesimal displacement of the path
changes the stacking sequence (except for the special case in which the displacement is such that the path moves along the boundary
of the stacking diagram). The asymptotic terminal lines that occur in the case of infinite sample height
do not appear at finite height since the slope of the sedimentation path is fixed and the paths are of finite length.

The three boundaries SBI, SBII, and TL divide the stacking diagram in different regions. Each region corresponds to a different stacking sequence. In order to identify each sequence we
first select one point inside of the desired region. Next we plot the corresponding path in the bulk phase diagram such that we can determine the sequence by inspecting the
crossings between the path and the binodals. 

Once the stacking diagram has been calculated in the $\bar\mu_1-\bar\mu_2$ plane, we can easily transform to any other set of variables provided that the equation of state 
of the mixture is known. In order to ease comparison to experimental work, a sensible choice of variables for the stacking diagram is
the $\bar\eta_1-\bar\eta_2$ plane, where $\bar\eta_i$ is the average packing fraction of species $i$,
\begin{equation}
\bar\eta_i=\frac1h\int_0^hdz\eta_i(z).
\end{equation}
Here, $\eta_i(z)$ is the local packing fraction of species $i$ at a distance $z$ from the bottom of the sample. To obtain $\eta_i(z)$ we 
first compute the local chemical potentials at $z$ using Eq.~\eqref{chemi}, and then use the equation of state of the system $\eta_i=\eta_i(\mu_1,\mu_2)$. The
phase diagram in the $\bar\eta_1-\bar\eta_2$ plane is then obtained by transforming the coordinates of the boundaries in the stacking diagram from $(\bar\mu_1,\bar\mu_2)$
to $(\bar\eta_1,\bar\eta_2)$. Other representations of the stacking diagram such as for example average osmotic pressure versus average composition are also possible,
following a similar transformation procedure.

\section{Results}

We first apply our theory to obtain the stacking diagrams at finite height of different bulk model phase diagrams (Sec.~\ref{aaa}). Although the bulk phase diagrams do not correspond
to real microscopic models, they are representative of the behaviour of typical colloidal mixtures. The model bulk phase diagrams provide relevant examples of possible
topologies of the stacking diagrams. 
In Sec.~\ref{bbb} we apply the theory to model binary
mixtures of patchy colloids for which we use Wertheim's theory to obtain the bulk phase diagram. Finally, in Sec.~\ref{ccc} we compare the stacking diagrams at finite and
infinite sample heights.

\subsection{Model bulk phase diagrams}\label{aaa}
\begin{figure*}
\epsfig{file=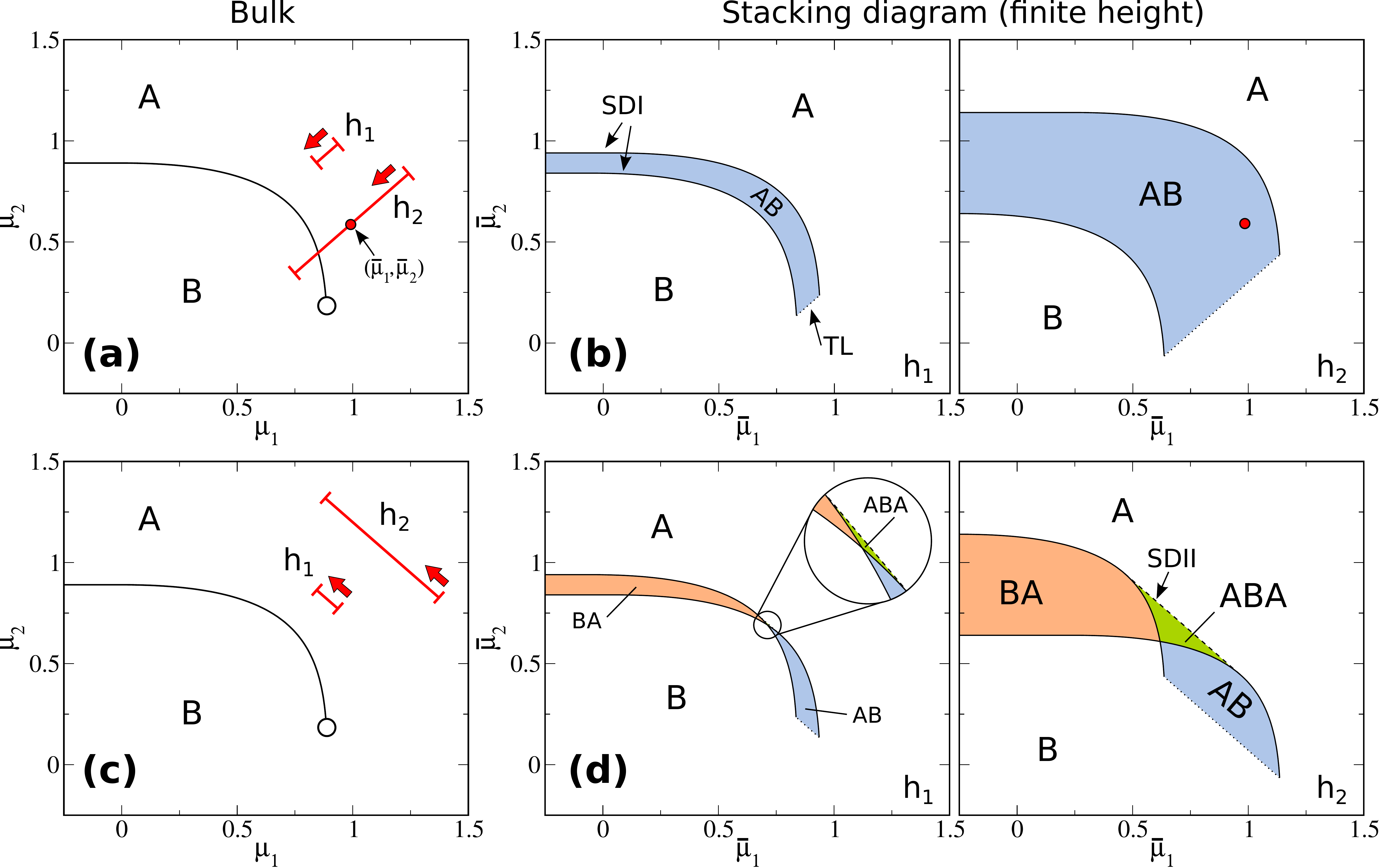,width=0.8\textwidth}
\caption{(a) Schematic bulk phase diagram of a binary mixture in the plane of chemical potentials $\mu_1-\mu_2$. Two phases A and B coexist at the binodal (solid black line). The binodal ends at 
a critical point (empty circle). The solid red lines are representative sedimentation paths (finite height) corresponding to samples with two different heights, $h_1$ and $h_2>h_1$, as indicated. The
slope of the path is $s=1$ in both cases. The arrow indicates the direction of the path from the bottom to the top of the sample. The red circle on the path $h_2$
is located at the center of the path. Its coordinates are the average local chemical potentials along the path. 
(b) Stacking diagrams for samples with heights $h_1$ (left) and
$h_2$ (right). Sedimentation binodals of type I (SBI) are indicated by black solid lines and terminal lines (TL) by black dotted lines. Each region is coloured and labelled according to its
stacking sequence (from bottom to top). The sedimentation path corresponding to the point highlighted with a red circle in panel (b) is represented in the bulk phase diagram (a).
The lower panels (c) and (d) show the same diagrams for the case of sedimentation paths with slope $s=-1$. The sedimentation binodals of type II are indicated
by black dashed lines. The inset on the left of panel (d) is a close view of a region of the stacking diagram.} 
\label{fig3}
\end{figure*}
One of the simplest possible bulk phase diagrams of a binary mixture is schematically represented in Fig.~\ref{fig3}a. There is a single binodal at which two phases
A and B coexist. The binodal ends at a critical point. The species $2$ undergoes an A-B phase transition. Hence, the binodal has a horizontal asymptote and tends to the
value of $\mu_2$ at the transition (chosen as $\lim_{\mu_2\rightarrow\infty}=0.89$). This phase diagram might be representative of e.g. a mixture of spherical colloids (species $1$)
and anisotropic colloids undergoing an isotropic-nematic phase transition (species $2$). A small degree of polydispersity in the spherical colloids could prevent a liquid-solid
phase transition in the pure system of species $1$~\footnote{This ignores phase  coexistence phenomena in polydisperse systems}.

The stacking diagram of the chosen model bulk phase diagram for the case of infinite height is shown in Fig. 5 of
Ref.~\cite{stack2}. Here we focus on the case of finite height. In Fig.~\ref{fig3}b we represent
the stacking diagram ($\bar\mu_1-\bar\mu_2$ plane) for two different heights $h_1$ and $h_2$ with $h_2>h_1$. In both cases the
slope of the paths are the same, $s=m_2/m_1=1$, and both buoyant masses are positive such that both local chemical
potentials decrease from the bottom to the top of the sample. We show representative sedimentation
paths in Fig.~\ref{fig3}a. Each sedimentation path in the $\mu_1-\mu_2$ plane in the bulk diagram is a point in the $\bar\mu_2-\bar\mu_1$ plane of the stacking diagram
(the coordinates of the middle point of the path).
The stacking diagrams contain two sedimentation binodals of type I (generated by paths starting and ending at the bulk binodal) 
and one terminal line (paths crossing the critical point). There are three possible stacking sequences, namely A, B, and AB. We label 
the sequences according to the order of different stacks from bottom to top.

Next, we study the same model bulk phase diagram but for sedimentation paths with a different slope, $s=-1$. See representative paths in Fig.~\ref{fig3}c.
Here $m_1>0$ and $m_2<0$ such that $\mu_1$ ($\mu_2$) decreases (increases) from the bottom to the top of the sample. The slope of the path is, in this case, 
compatible with the slope of the binodal in the sense that there is one point at the binodal whose derivative equals the slope of the path, cf.~\eqref{mmm}. Hence,
the stacking diagram contains a sedimentation binodal of type II which is formed by the set of paths that are tangent to the binodal in bulk. The boundaries of the stacking diagrams  
(Fig.~\ref{fig3}d) are: two SDI lines, one SDII line, and one TL. These boundaries split the stacking diagram into five regions. The possible stacking sequences are
A, B, AB, ABA, and BA~\footnote{Note that sequences with a single stack, such as A, are actually one phase-systems and not proper sequences made of different stacks}. The ABA sequence appears when a path crosses the bulk binodal twice~\cite{floating,schmidt04aog}.

This very simple example already shows the richness of the stacking diagram. It also suggests that the sample height 
plays a major role. The size of the area of the stacking diagrams occupied by each stacking sequence depends strongly on the height of the sample. For example,
the AB region substantially increases with $h$, cf. Fig.~\ref{fig3}b.
Two samples of different height and different stacking sequences might have the same composition and concentration of colloids (we will
see examples in the next section). The height of the sample might have an even stronger influence on the stacking diagram, as we will demonstrate in the following.

\begin{figure*}
\epsfig{file=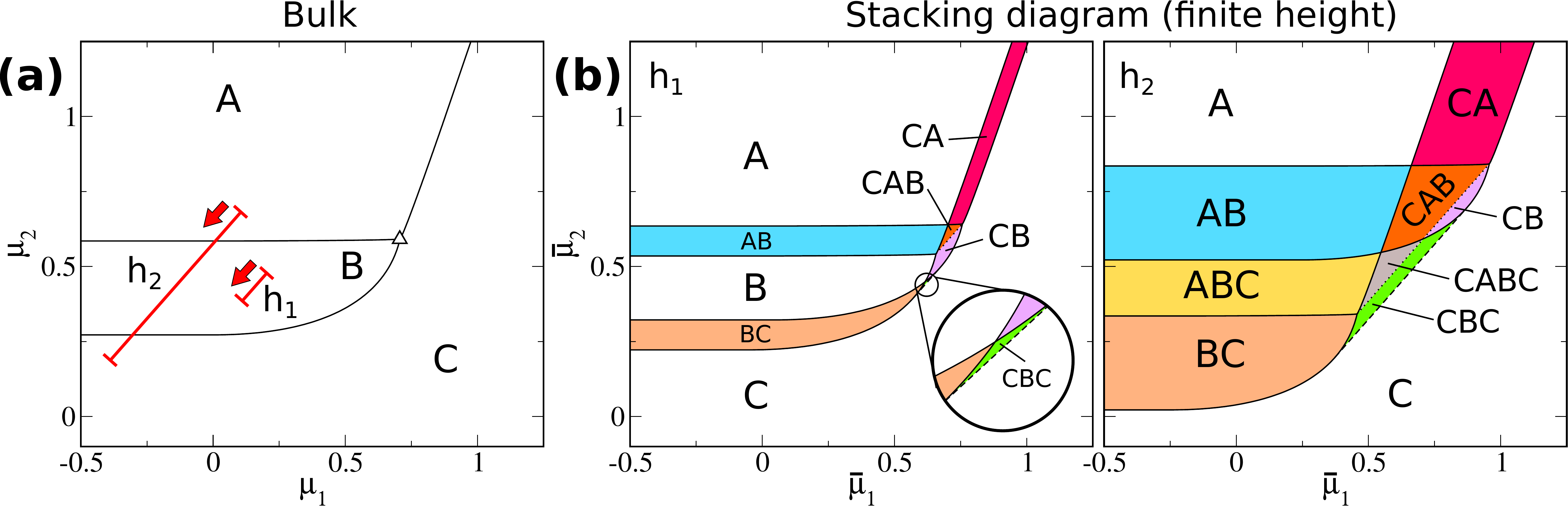,width=0.9\textwidth}
\caption{(a) Schematic bulk phase diagram of a binary mixture in the plane of chemical potentials $\mu_1-\mu_2$. The solid black lines indicate binodals. Three phases A, B, C coexist at a triple point
(triangle). The solid red lines are representative sedimentation paths of samples with two different finite heights, $h_1$ and $h_2$, as indicated. The
slope of each path is $s=1$. The arrow indicates the direction of the path from the bottom to the top of the sample. 
(b) Stacking diagrams for samples with heights $h_1$ (left) and
$h_2>h_1$ (right). The SBI lines are indicated by black solid lines. The SBII lines are represented as black dashed lines. The TL are indicated by black dotted lines. 
Each region is coloured and labelled according to its
stacking sequence (from bottom to top). 
The inset on the left of panel (b) is a close view of a region of the stacking diagram.} 
\label{fig4}
\end{figure*}

In Fig.~\ref{fig4}a we show a further model bulk phase diagram. There are three different phases: A, B and C. Three binodals for A-B, A-C, and B-C coexistence
meet at a triple point. A phase 
diagram like this  might represent a mixture in which the species $1$ represents spherical colloids and the species $2$ consists of e.g., elongated colloidal particles. 
The elongated particles can undergo isotropic-nematic and nematic-smectic phase transitions. 

The stacking diagrams for this mixture are depicted in Fig.~\ref{fig4}b for two different heights, $h_1$ and $h_2$, with $h_1<h_2$. In both cases the slope
of the path is $s=1$ and both buoyant masses are positive. The boundaries in the stacking diagram are: six SDI lines (two for each of the three binodals),
one SDII line (the
slope of the path matches the slope of the B-C binodal at one point), and one TL line (originating from the triple point). The stacking diagrams
for heights $h_1$ and $h_2$ differ substantially from each other, see left and right panels of Fig.~\ref{fig4}b, respectively. We observe two main differences between the diagrams
for short and long samples:

First, the sedimentation paths for the small system ($h_1$) fit in the space between the A-B and B-C binodals of the bulk phase
diagram, see an example in Fig.~\ref{fig4}a. Consequently the stacking sequence B occurs in the stacking diagram, Fig.~\ref{fig4}b (left). In contrast, the
stacking sequence B does not occur in the large samples ($h_2$). The B sequence is replaced by an ABC state, Fig.~\ref{fig4}b (right). The sedimentation paths
in this case are long enough such that they do not fit in the region between the A-B and B-C binodals in bulk. Instead, the path must cross at least one of the binodals.

Second, the sequence CABC is generated by paths crossing the three binodals in bulk. This sequence is present only in the long samples, Fig.~\ref{fig4}b (right). 
The paths corresponding to the short samples ($h_1$) are not long enough to cross the three binodals, and hence the CABC sequence is absent.

These examples illustrate how the stacking diagrams for different heights might differ qualitatively. By changing the overall height of the sample some stacking sequences
are replaced by others (e.g. the B sequence for $h_1$ is replaced by ABC for $h_2$) and one also observes the occurrence of new sequences, such as the CABC sequence for $h_2$.

\subsection{Mixtures of patchy colloids}\label{bbb}
We next apply our theory to patchy colloidal binary mixtures. 
We study two cases in which the species differ either by the number or by the types of patches.

\begin{figure}
\epsfig{file=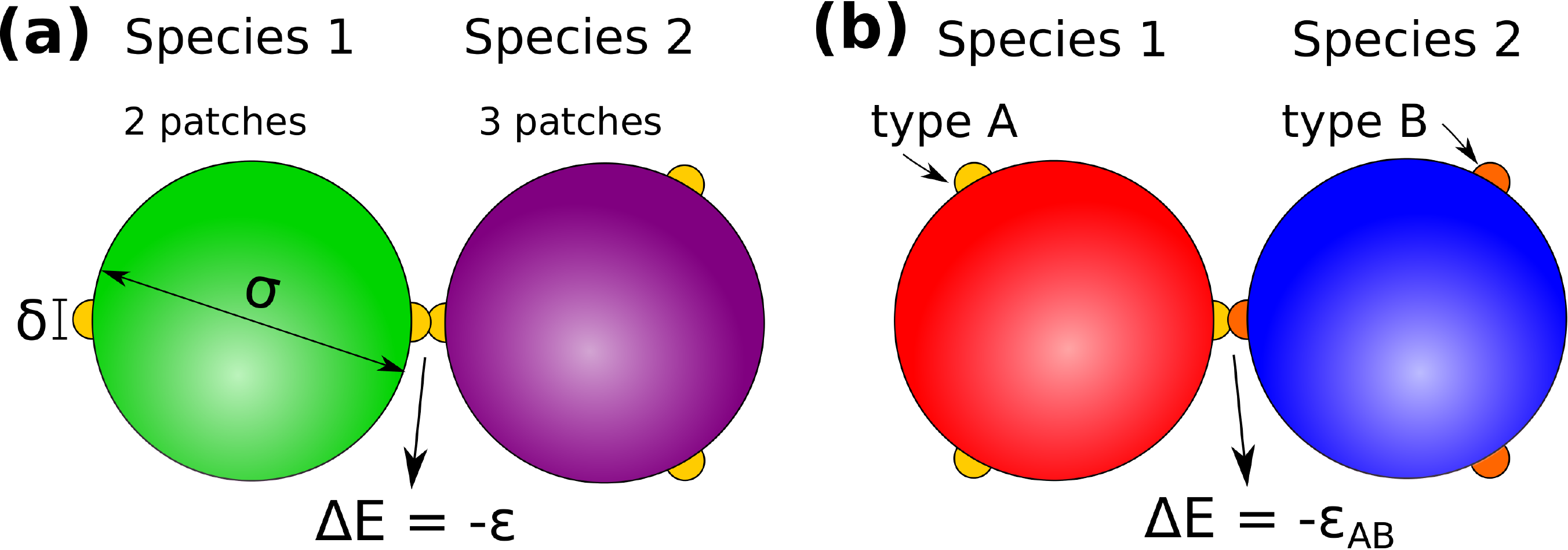,width=0.95\columnwidth}
\caption{Schematic of the patchy colloidal particles. All colloids are modelled as hard spheres of diameter $\sigma$ with patches on the surface (spheres of diameter $\delta\approx0.12\sigma$).
We study two types of mixtures. (a) Binary mixtures of particles with identical patches. The species $1$ has $2$ patches and the species $2$ has $3$ patches. (b) Binary mixture of
particles with three patches of type A (species 1) and of type B (species 2). If two patches overlap the internal energy of the system decreases in a quantity given by the type of
patches involved.}
\label{fig5}
\end{figure}

\subsubsection{Different number of patches}\label{23m}
We model the colloids by hard spheres of diameter $\sigma$ with identical patches (spheres of size $\delta$) on the surface, see Fig.~\ref{fig5}a. If two patches overlap the internal energy of the system
decreases by $\epsilon$. We use Wertheim's first order perturbation theory~\cite{wertheim} and a generalization of the Flory-Stockmayer theory of polymerization~\cite{flory,stock}
to compute the bulk phase diagram of the mixture. We follow exactly the same implementation of the theory as in Ref.~\cite{C0SM01493A}. Theory and Monte Carlo simulations for the bulk phase behaviour
are in semi-quantitative agreement with each other~\cite{emptyscio,felix}. 

\begin{figure*}
\epsfig{file=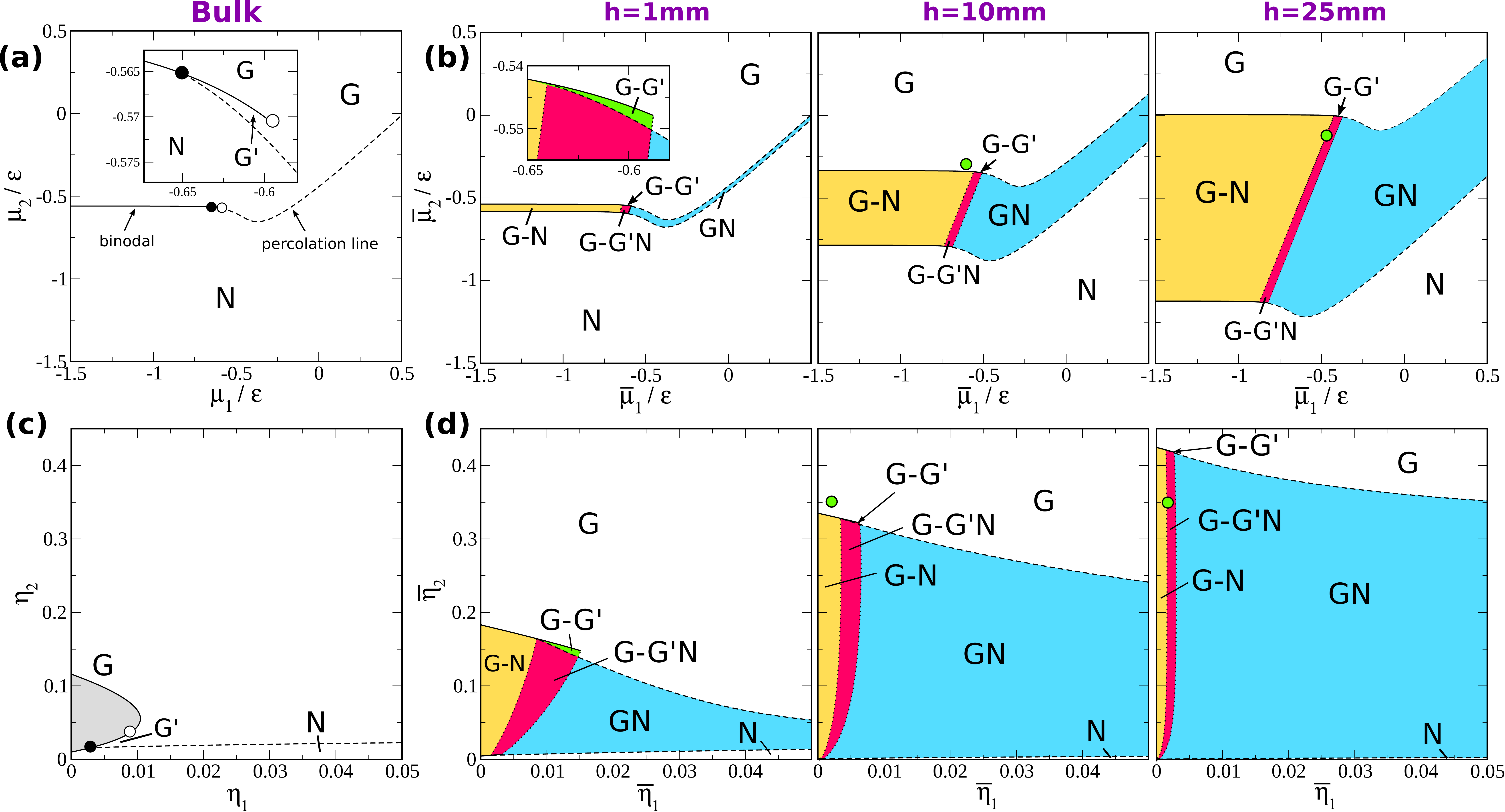,width=0.95\textwidth}
\caption{Bulk phase diagram of a binary mixture of patchy colloids with two (species 1) and three (species 2) identical patches in the plane of chemical potentials
(a) and packing fractions (c). The temperature is $k_{\text{B}}T/\epsilon=0.09$.
The black solid line indicates the binodal. The dashed line is the percolation lines of the mixture.
The empty circle indicates the critical point. The black solid circle is the ending point of the percolation line. The  inset in (a) is a close view of the region
close to the critical point.
(b) and (d) show the stacking diagrams of the mixture under gravity in the plane
of average chemical potentials (b) and average packing fractions (d) for three different sample heights, as indicated. A dash
between two letters, like in the sequence G-N, indicates that the sedimentation path crosses the binodal. The absence of a dash, such as e.g. in GN, indicates that
the path crosses the percolation line.}
\label{fig8}
\label{fig6}
\end{figure*}

The species $1$ has two patches, and the species $2$ has three patches.
The colloids with three patches undergo a phase transition between two fluids with different densities.
With only two patches present the particles of species $1$ can form only chains. The absence of branching prevents 
phase separation and there is no fluid-fluid phase transition in the pure system of species $1$.  In the mixture the 
transition between high and low density fluids ends at a critical point.
See the binodal in the bulk phase diagram of the mixture shown in Fig.~\ref{fig6}a for 
scaled temperature $k_{\text{B}}T/\epsilon=0.09$.
In addition to the binodal, the phase diagram contains a percolation line that divides 
percolated and non-percolated states. The system is percolated if the probability that a patch is bonded, $f_{\text{b}}$, is
higher than the percolation threshold $p_{\text{T}}$. The percolation line intersects the  binodal close to the critical point on the low density side.
The high density phase (G) is an equilibrium gel or network fluid which is always percolated. The low density phase does not percolate (N) except for a very
narrow region close to the critical point (G'). We refer the reader to Refs.~\cite{emptyscio,C0SM01493A} for further details about the bulk phase behaviour of this mixture.

To proceed and to obtain the stacking diagrams we need to set the slope of the sedimentation paths and the height of the sample.
We fix the gravitational lengths of the colloids to $\xi_1=5\text{ mm}$ and $\xi_2=2\text{ mm}$ (typical values for 
colloidal particles). Hence, the slope of the path is fixed to $s=\xi_1/\xi_2=2.5$. The stacking diagrams in the $\bar\mu_1-\bar\mu_2$ plane for
three different heights $h=1$ mm, $10$ mm, and $25$ mm are shown in Fig.~\ref{fig6}b. Each of them contains four SDI lines (two for
the binodal and two for the percolation line) and two terminal lines (one for the critical point and one for the ending point
of the percolation line). Six different stacking sequences are possible for this value of the slope: \mbox{N, G, GN, G-N, G-G', and G-G'N.} We use a dash between two stacks 
in the stacking sequence, like in the G-N sequence, to indicate that the sedimentation path crosses the binodal. The absence of a dash, e.g. in the GN sequence, indicates that
the path crosses the percolation line. 

Once the stacking diagrams in the plane of average chemical potentials have been computed, we can transform the 
variables using the procedure described at the end of Section~\ref{theory}. In Fig.~\ref{fig6}d we show the resulting stacking
diagrams in the plane of average packing fractions. 

\begin{figure}
\epsfig{file=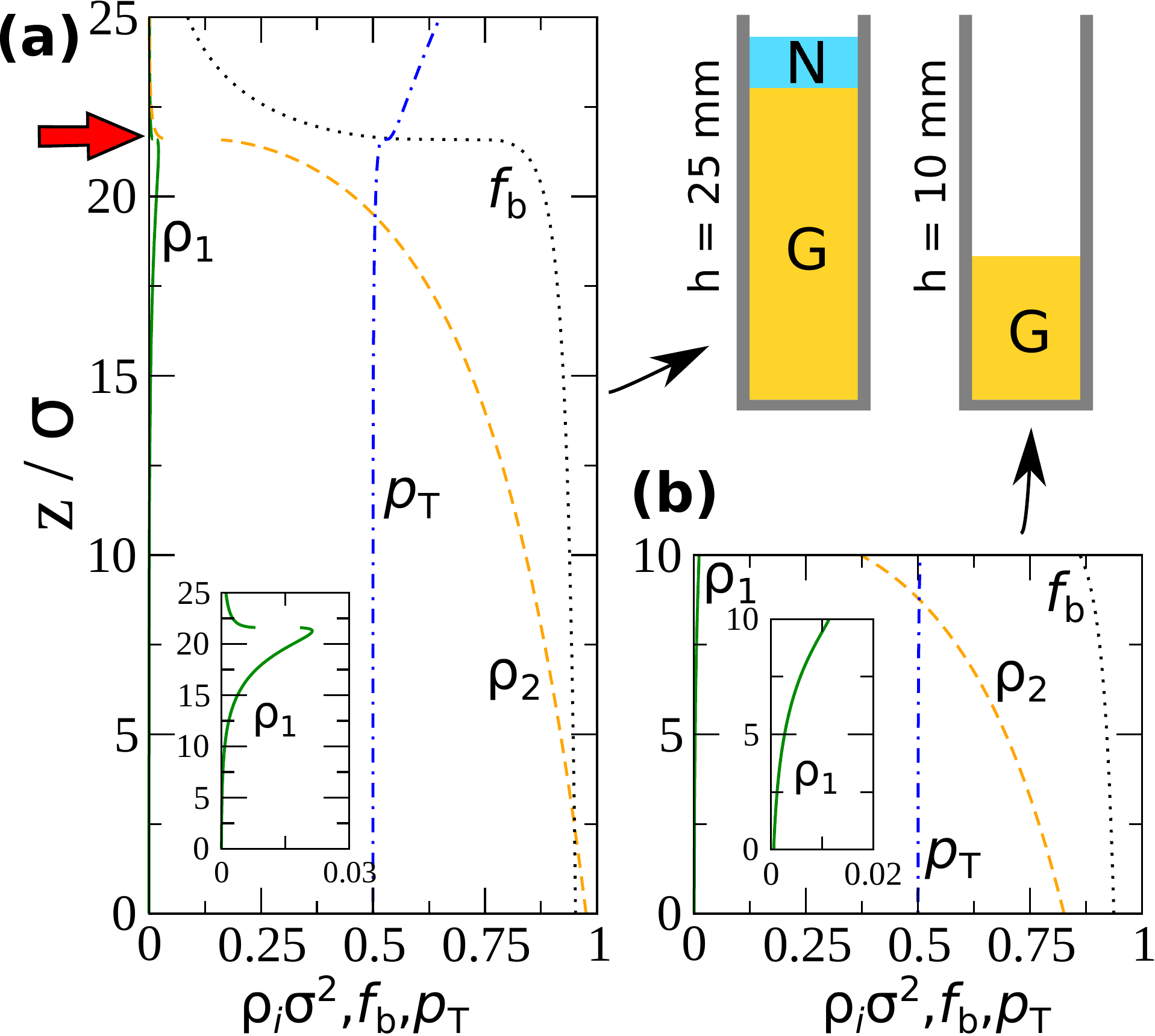,width=0.9\columnwidth}
\caption{Density profiles $\rho_i$ of species 1 (green solid line) and species 2 (orange dashed line), percolation threshold $p_\text{T}$ (dot-dashed blue line),
and bonding probability $f_{\text{b}}$ (dotted black line) of a binary mixture of patchy particles with two (species $1$) and three (species 2) patches under gravity.
The height of the container is $h=25$ mm (a) and $10$ mm (b). The system is percolated if $f_{\text{b}}>p_{\text{T}}$.
In both cases the average packing fractions of the colloids of each species are the same,
$\bar\eta_1=0.002$, and
$\bar\eta_2=0.35$. 
The stacking sequences are G-N (a) and G (b), schematically represented in the upper right corner of the figure. 
The insets in (a) and (b) are close views of the $\rho_1$ profile. The red arrow in panel (a) indicates the position of the G-N interface.}
\label{fig7}
\end{figure}

The number and types of stacking sequences remain the same for the
sample heights investigated here. However, the region of the phase space occupied by each sequence significantly depends
on the value of the sample height. We show a specific example in Fig.~\ref{fig7} in which we plot the density profiles of two samples
with the same average packing fractions ($\bar\eta_1=0.002$, and $\bar\eta_2=0.35$), but different heights ($h=25$ mm and $10$ mm).
The corresponding state points are highlighted by green solid circles in the stacking diagrams of Fig.~\ref{fig6}. Despite the average
colloidal concentrations being the same, the stacking sequences differ: G-N for the sample with $h=25$ mm and G for the case $h=10$ mm.
Other values of the sample height and the gravitational lengths will result in identical phenomenology provided that the ratios $h/\xi_i$ with
$i=1,2$, are unchanged.

\subsubsection{Different types of patches}
As a concluding example we study a binary mixture of patchy colloids with different types of patches. The species $1$ ($2$) possesses three patches of type A (B), see Fig.~\ref{fig5}b for an illustration.
When two patches of type $\alpha$ and $\beta$ with $\alpha,\beta=\{A,B\}$ overlap, the energy of the system decreases by $\epsilon_{\alpha\beta}$. The bulk
phase diagram of this model has been studied theoretically~\cite{C1SM06948A} and by Monte Carlo simulations~\cite{felix} for different values of the bonding energies $\epsilon_{\alpha\beta}$. The phenomenology
that emerges is very rich as different types of gels can occur depending on the set of bonding energies. Here, we set
$\epsilon_{\text{BB}}=\epsilon$ (energy scale), $\epsilon_{\text{AA}}=0.80\epsilon$, and $\epsilon_{\text{AB}}=0.85\epsilon$.
We fix the scaled temperature, as in the previous case, to $k_{\text{B}}T/\epsilon=0.09$. 

The bulk phase diagrams in the planes of chemical potentials and of packing fractions are shown in Fig.~\ref{fig8}a and ~\ref{fig8}b, respectively. At the value of temperature considered
only the species $2$ (strongest bonds)
undergoes a fluid-fluid phase transition. Hence, in the mixture there is only one binodal that ends at a critical point. 
In addition there are three percolation lines. One of these indicates whether the full
mixture percolates, and the other two percolation lines indicate whether the individual species percolate. 
Although species $1$ does not undergo a first order fluid-fluid phase transition at this temperature, it still undergoes a percolation transition.
The percolation lines and the binodal divide the bulk phase diagram into five different regions. At low chemical potentials (and hence low densities) the system is non-percolated (N). The other four states are equilibrium percolated gels: (i) a mixed gel (M) in which the mixture
percolates but none of the species percolates independently, (ii) a bicontinuous gel or bigel (B) in which the mixture and both species percolate, (iii) two gels (G$_i$, $i=1,2$) in which
the mixture and the species $i$ percolate. See~\cite{C1SM06948A,felix} for further details about the bulk behaviour. 

\begin{figure*}
\epsfig{file=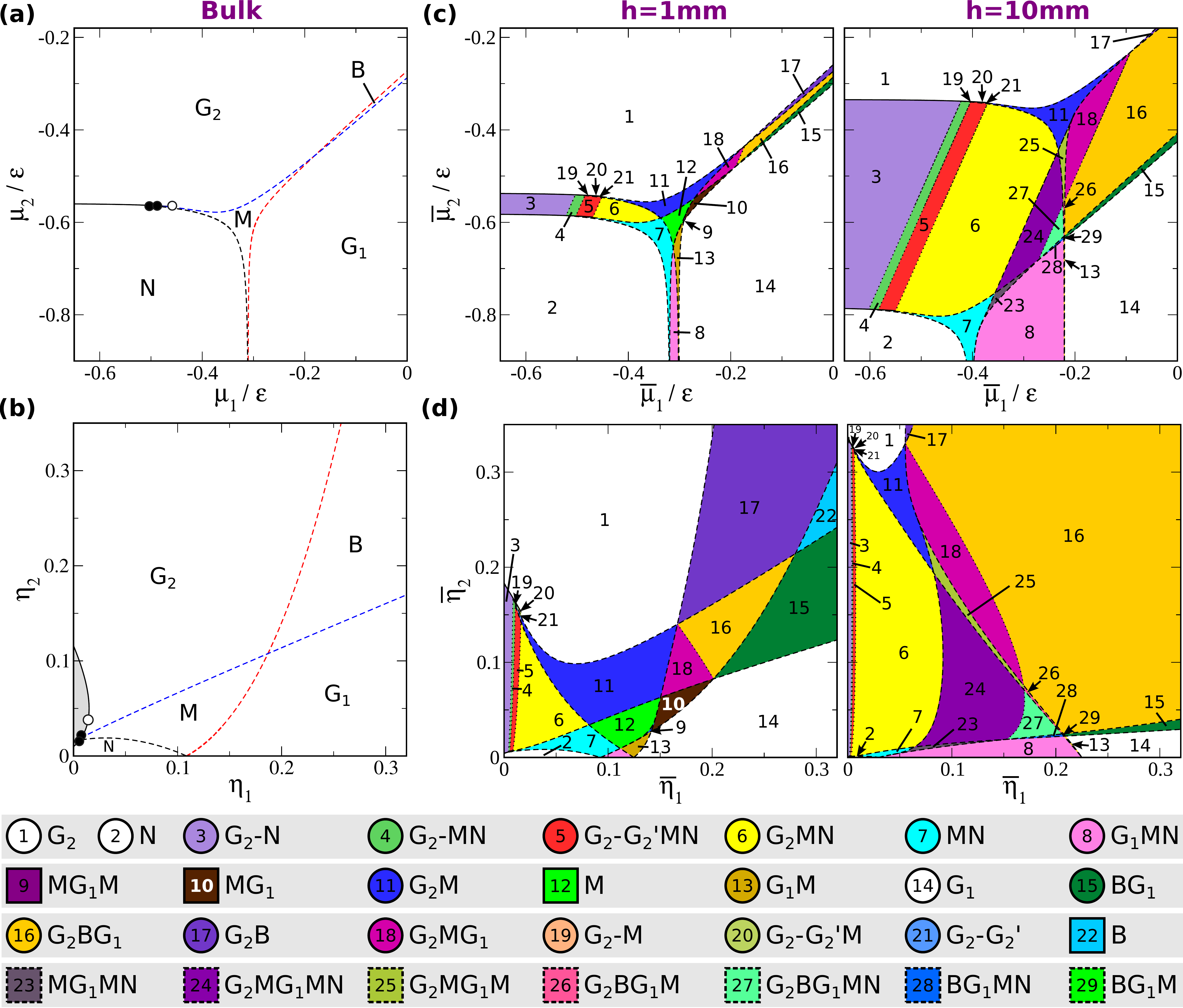,width=0.80\textwidth}
\caption{Bulk phase diagram of a binary mixture of patchy colloids with three patches of type A (species 1) and three patches of type B (species 2) in the plane of chemical potentials
(a) and packing fractions (b). The black solid line indicates the binodal. The dashed lines are percolation lines of the mixture (black), of the species 1 (red), and of the species 2 (blue).
The empty circle indicates the critical point. The black solid circles are the ending points of the percolation lines. Shown are the stacking diagrams of the mixture under gravity in the plane
of average chemical potentials (c) and average packing fractions (d) for two different sample heights, as indicated. The legend shows the occurring stacking sequences. A dash
between two letters, like in the sequence 3, G$_2-$N indicates that the sedimentation path crosses the binodal. The absence of a dash, such as e.g. in 13 (G$_2$M) indicates that
the path crosses a percolation line. The sequences marked  with a circle in the legend are present at both sample heights. The sequences marked with a solid (dashed) square occur
only for samples with $h=1$ mm ($h=10$ mm). 
}
\label{fig8}
\end{figure*}

Here, we study sedimentation-diffusion-equilibrium. As in the example above we chose the gravitational lengths to be $\xi_1=5$ mm and $\xi_2=2$ mm, and study two different
sample heights $h=1$ mm and $10$ mm. The resulting stacking diagrams are extremely rich, see Fig.~\ref{fig8}c and d, with more than $20$ distinct stacking sequences. Again, the regions
occupied for each stacking sequence depend on the sample height. In some cases the same stacking sequence occurs in a completely different range of average packing fractions
when varying the sample height, see for example the sequence 11 (G$_2$M) in Fig.~\ref{fig8}d. Even more important is the fact that the stacking diagrams for $h=1$ mm and $h=10$ mm are
{\it qualitatively} different. There are several stacking sequences that are present only in one of the selected sample heights. For example, the sequence MG$_1$ (number $10$
in Fig.~\ref{fig8}) is only present in samples with $h=1$ mm, and the sequence G$_2$MG$_1$M (number $25$) occurs only for the case $h=10$ mm.

\subsection{Infinite vs finite height stacking diagrams}\label{ccc}
\begin{figure}
\epsfig{file=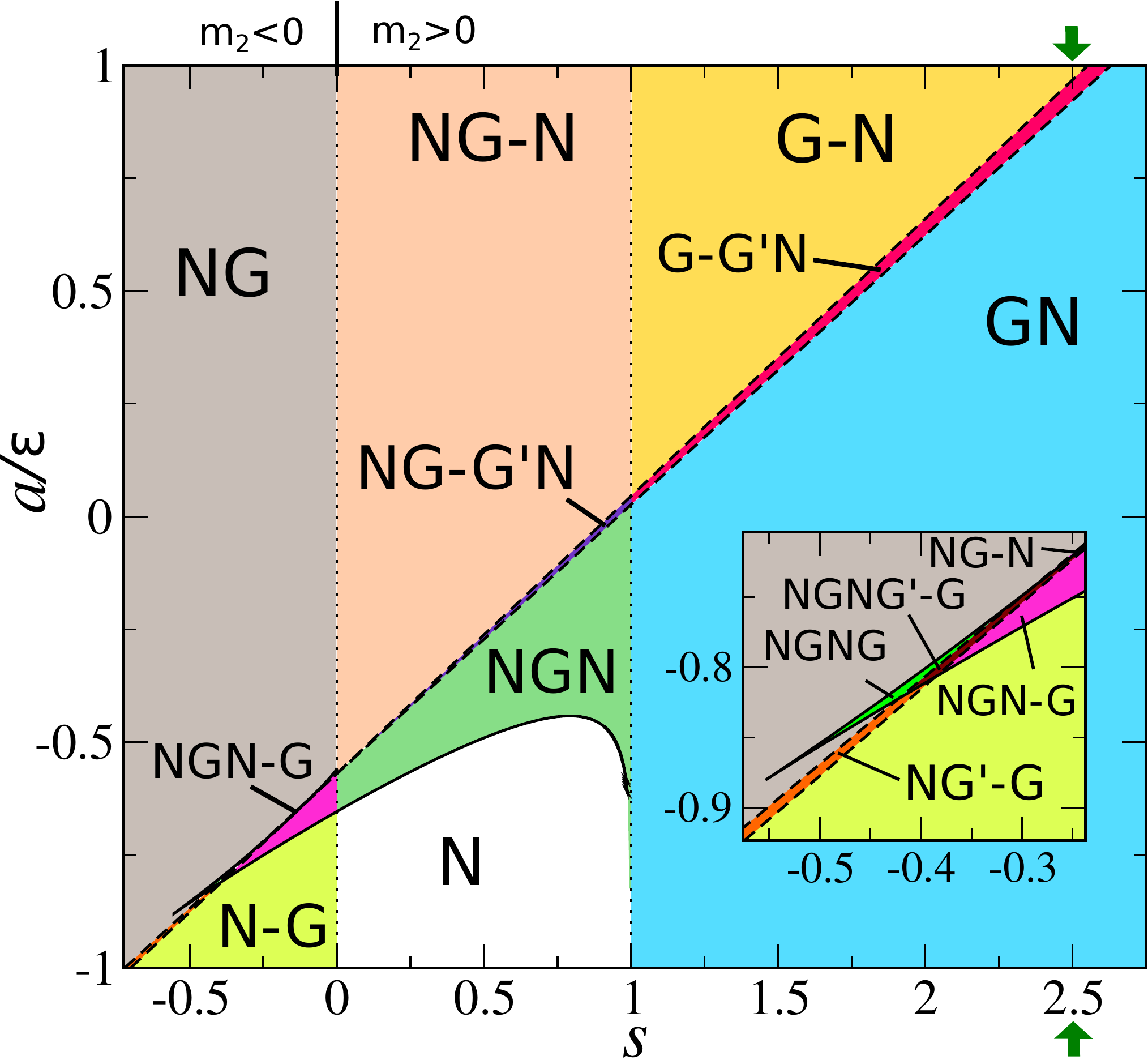,width=0.80\columnwidth}
\caption{Stacking diagram (infinite height) in the $s-a$ plane of a binary mixture of patchy colloids with two and three patches at $k_{\text{B}}T/\epsilon=0.09$. The inset
is a close view of a small region. Each coloured region is a different stacking sequence labelled from bottom to top of the sample. A dash in the label indicates the
sedimentation path crosses the binodal. No dash indicates the path crosses a percolation line. The stacking diagram corresponds to the case $m_1>0$. The stacking 
diagram for the case $m_1<0$ can be obtained by just reversing the order of the stacks. The green arrows signal the slope $s=2.5$ for which we have obtained 
the stacking diagrams at finite height, see Fig.~\ref{fig6}. Sedimentation binodals are represented by solid lines. Terminal lines are shown as dashed lines. Asymptotic
terminal lines are indicated by dotted lines.}
\label{fig9}
\end{figure}
We conclude the section with several comments regarding the connection between the stacking diagrams for infinite and finite samples. The main effect is the occurrence
of new sequences in the case of finite height samples. The new sequences are formed by the removal of one or more stacks of the sequences for $h\rightarrow\infty$. In
general, a sequence observed at finite height might be a truncated sequence of the infinite system. This observation has strong implications for the correct interpretation of observed
stacking sequences in finite height samples. 

In Fig.~\ref{fig9} we show the stacking diagram (infinite height) of the mixture of patchy colloids with two and three patches analyzed in Sec.~\ref{23m}. The
stacking diagram is represented in the $s-a$ plane, cf~\eqref{constants}. The diagram has been computed for $m_1>0$. Hence positive (negative) values of the slope $s=m_2/m_1$
correspond to positive (negative) values of $m_2$. There exists an analogous diagram for $m_1<0$  in which the only difference is that the stacking sequences
have the reverse order. There are two sedimentation binodals (one for the binodal and one for the
percolation line), two terminal lines (critical point of the binodal and ending point of the percolation point), and two asymptotic terminal lines (asymptotic behaviour
of the binodal and the percolation lines). 

We have shown previously the diagrams at finite height for the slope $s=2.5$, see Fig.~\ref{fig6}.
For this value of $s$ only three sequences are possible at infinite height, see Fig.~\ref{fig9}: \mbox{GN, G-G'N, and G-N}. The finite height 
diagram is richer with up to six different sequences. These are the same three as for $h\rightarrow\infty$ and three new truncated sequences of the infinite 
case (G,N, and G-G'). As expected, by increasing the height of the sample, the regions occupied by the new truncated sequences in the stacking diagrams shrink (see Fig.~\ref{fig6}d).
In this particular example, for $h=25$ mm, the stacking diagram is already dominated by the stacking sequences of the infinite height case.

The infinite height stacking diagram provides the set of possible sequences for different values of $s$. Here, 
we have only analysed the value $s=2.5$ of the slope for finite height samples. The infinite height stacking diagram shows that for other values of $s$
further complex phenomenology occurs. For example, for negative values of the slope, i.e. $m_2<0$, it is possible to stabilize the sequence NGN-G
which constitutes a reentrant percolation phenomena. This sequence also occurs in two-dimensional binary mixtures of patchy colloids~\cite{Lucas}.

Experimentally one can change the slope of the path
via the synthesis of colloids with cores of different materials~\cite{doi:10.1021/la0101548,doi:10.1021/am400490h} or changing the mass density of the solvent.
Hence, the full range of stacking sequences of a given colloidal mixture is, in principle, experimentally accessible.

\section{Discussion and conclusions}

Our theory is based on a local density approximation which assumes that for each  $z$ the state of the sample can be approximated by a bulk state. 
Non-local effects might modify the stacking diagrams. 
In particular, the theory neglects the surface tension of the interfaces between stacks in the stacking sequence.
If one of the stacks is very narrow the surface tension of the upper and lower interfaces might be higher than the gain in free energy due to the formation of the stack,
as observed in colloid-polymer model mixtures~\cite{schmidt04aog}. 
Under such circumstances, the final equilibrium stacking sequence might be different than that predicted by our local theory. This condition is analogue of capillary
condensation/evaporation.
Surface effects such as e.g., the occurrence of wetting and layering near the walls of the vessel might also modify the stacking diagrams. 

Our theory can be easily extended to multicomponent systems since the sedimentation paths remain lines in the phase space of chemical potentials. Also, the
theory is directly applicable to molecular systems. There, the gravitational lengths are orders of magnitude higher than in colloidal systems. Hence, to observe
similar phenomenology one needs containers of considerable size, such as for example geological deposits. 

We have obtained the stacking diagrams at constant sample height and fixed ratio of the buoyant masses. Other choices, such as for example keeping the colloidal
concentrations fixed and varying the sample height, are also possible. A stacking diagram in which one of the variables is the height might be relevant to study
the effects of slow solvent evaporation, which is a process that changes the total volume but keeps the particle number fixed.

We have shown that two samples with the same colloidal concentrations
but placed in vessels of different heights might have different stacking
sequences. We have also shown that the stacking diagrams might be qualitatively different for different heights. Therefore, the sample height plays a major role in sedimentation-diffusion-equilibrium experiments. This role is as important as for example the average colloidal concentrations. We conclude that the sample height should be carefully measured and specified in any sedimentation
experiment. 

We have focused on sedimentation-diffusion-equilibrium in colloidal mixtures. 
Future studies could consider the dynamics of sedimentation using 
dynamic density functional theory~\cite{evans_dft,ddft} and power functional theory~\cite{pft}. 

\begin{acknowledgments}
We thank Matthias Schmidt for very useful discussions and a careful reading of the manuscript.
This work was supported in part by the Portuguese Foundation for Science and Technology (FCT) through project
EXCL/FIS-NAN/0083/2012.
\end{acknowledgments}

\end{document}